\documentclass[aps,pre,showpacs,twocolumn,superscriptaddress,floatfix]{revtex4} 
\usepackage{graphicx} 

\begin{document} 
 
\preprint{version 21} 

\title{A simulational and theoretical study of the spherical electrical double layer for a 
size-asymmetric electrolyte: the case of big coions.
} 

\author{G. Iv\'an Guerrero-Garc\'{\i}a}
\affiliation{Instituto de F\'{\i}sica, Universidad Aut\'onoma de San Luis Potos\'{\i}, \\
\'Alvaro Obreg\'on 64, 78000  San Luis Potos\'{\i}, San Luis Potos\'{\i}, M\'exico} 

\author{Enrique Gonz\'alez-Tovar} 
\affiliation{Instituto de F\'{\i}sica, Universidad Aut\'onoma de San Luis Potos\'{\i}, \\
\'Alvaro Obreg\'on 64, 78000  San Luis Potos\'{\i}, San Luis Potos\'{\i}, M\'exico} 

\author{Mart\'{\i}n Ch\'avez-P\'aez}
\affiliation{Instituto de F\'{\i}sica, Universidad Aut\'onoma de San Luis Potos\'{\i}, \\
\'Alvaro Obreg\'on 64, 78000  San Luis Potos\'{\i}, San Luis Potos\'{\i}, M\'exico}

\date{\today}  
 
\begin{abstract} 

Monte Carlo simulations of a spherical macroion, surrounded by a size-asymmetric
electrolyte in the primitive model, were performed. We considered 1:1 and 2:2 salts
with a size ratio of 2 (i.e., with coions twice the size of counterions),
for several surface charge densities of the macrosphere. The radial distribution functions, 
electrostatic potential at the Helmholtz surfaces, and integrated charge  
are reported. We compare these simulational data with original results obtained from the
Ornstein-Zernike integral equation, supplemented by the hypernetted chain/hypernetted chain (HNC/HNC)
and hypernetted chain/mean spherical approximation (HNC/MSA) closures, and with
the corresponding calculations using the modified Gouy-Chapman and unequal-radius
modified Gouy-Chapman theories. The HNC/HNC and HNC/MSA integral equations formalisms 
show good concordance 
with Monte Carlo ``experiments'', whereas the notable limitations of point-ion approaches
are evidenced. Most importantly, the simulations confirm our previous theoretical
predictions of the non-dominance of the counterions in the size-asymmetric spherical
electrical double layer [J. Chem. Phys. {\bf 123}, 034703 (2005)], the appearance 
of anomalous curvatures at the outer Helmholtz plane and the enhancement 
of charge reversal and screening at high colloidal surface charge densities 
due to the ionic size asymmetry. 

\end{abstract}  
 
  
\maketitle 
  
\section{INTRODUCTION}

The study of charged colloidal solutions is very relevant for both basic research
and technology due to the ubiquitous nature of these systems  
\cite{Du01,Hu01,Vo01,Hu02,Fe01,Hi01}. Accordingly, the 
attainment of a successful theoretical description of such state of matter should 
represent a keystone for later developments in colloid science. For many years, 
the scientific community has investigated the structural characteristics of 
these materials, trying to understand the role of the electrostatic and 
entropic correlations in their observable properties. In particular, 
the interest in charged suspensions has prompted the burgeoning 
of unprecedented experimental techniques and of numeric and statistical mechanics 
approaches of increasing complexity. On the theoretical side, and despite
of the notorious progress in the speed of machine calculations, at present, it is 
not yet possible to mimic a real dispersion without making several and important 
simplifications in order to establish a tractable problem. Thus, for example, 
one of the most 
elemental idealizations of a diluted charged colloidal suspension is the combination 
of the cell and primitive models. Within this scheme, the average distance between 
non-concentrated macroions bathed by an electrolyte is very 
large, and therefore it is 
expected that the thermodynamics of the system will depend mainly on the 
ionic structure, or electrical double layer (EDL), around a
single macroparticle enclosed in an electroneutral cell. Complementarily, 
the so-called primitive model (PM), in which the ions are treated as
hard spheres with punctual charges embedded in their centers and the
solvent is considered a continuous medium, stands as the most thriving representation
of a multi-component electrolyte. A particular case of the PM is the restricted
primitive model (RPM), where all the ionic species are of equal size. This  
condition drastically facilitates the theoretical analysis and, as a consequence, 
a great amount of work has been performed in the RPM for the 
planar \cite{Ca01,Ch01,Ch02,Ch03,Bh001,Ji02},    
cylindrical \cite{Go01,Ji03,Goe02,Goe03}   
and spherical \cite{Go02,Outh01,Mes01,Mes02,Go03,Yu01,Goe01} double layers. In strong contrast, 
there are few articles in
which the effects of ionic size asymmetry have been studied systematically, and
these publications focus chiefly on the planar 
instance \cite{Va01,Bh01,Sp01,Kh01,Ma01,Ou01,Gr02,Val01,Gi01,Gue01,Ke01,Val02,Ke02}.

Certainly, the widespread use of the RPM to examine the double layer has led to
significant advances in the field, mainly due to its ability to explain a large
variety of colloidal phenomena, and, therefore, has established
it as the standard representation of the EDL. In turn, this adequacy
suggests that the RPM already contains most of the fundamental traits of a
colloidal suspension at the usual conditions of experiments and
applications. However, we consider that the lack of interest in
upgrading the model of a double layer so as to incorporate the effect
of ionic size asymmetry stems not only from the operational advantages
and/or from the past success of the RPM but also has been influenced
by the common belief in the {\it dominance of the counterions} in the EDL.
Such credence has its probable origin in a pioneering Poisson-Boltzmann (PB) study
by Valleau and Torrie \cite{Va01},
where they stated the following: ``...we expect the double layer
properties of a dilute (size-asymmetric) electrolyte to become similar to
those of a completely symmetric electrolyte having an effective size
equal to that of the counterion. (This remark will be asymptotically
exact for large fields in the Poisson-Boltzmann theory)...''. To be more
explicit, in Ref. \cite{Va01} a size-asymmetric electrolyte
next to an electrified wall was analyzed via a {\it quasi point-like ions} theory known
as unequal-radius modified Gouy-Chapman (URMGC), which in essence is equivalent
to the classical nonlinear PB equation for a binary mixture of punctual ions 
but with the assignment of different distances of closest approach (with respect to
the plate) to anions and cations. Therefore, the remark of counterion dominance,
quoted above, was indeed formulated and proved strictly
at the Poisson-Boltzmann level.

Notably, during the past years,
a great deal of modern treatments of the EDL
have also endorsed the
idea that counterions control the properties of high charged surfaces
immersed in both
size-symmetric \cite{Ca02,Yu01,Bh02} and size-asymmetric \cite{Kh01,Ma01,Val01,Gi01} 
primitive model 
(RPM and PM, respectively) electrolytes. Such agreement, about the dominance of 
counterions, between the current EDL theories and the old URMGC picture 
is remarkable since in all these new integral equations \cite{Ca02,Kh01,Ma01,Val01}, 
density functional \cite{Yu01,Gi01} and mean 
electrostatic potential \cite{Bh02} 
papers the fundamental hypothesis
of point-ions has been surpassed by including explicitly the
hard-core and electrostatic
correlations neglected in the PB theory. 
Notwithstanding, it must be noted that
in the cited beyond-PB surveys
the ``confirmation'' of the leading role of counterions has been based
in studies of either charge-asymmetric RPM electrical double layers
at low/moderate surface charges  \cite{Ca02} or, else, of size-asymmetric
systems near the point of zero charge (PZC) \cite{Ca02,Kh01,Val01}. In other words, therein
the original conclusion of the preponderance of counterions in the
EDL at {\it high} electric fields has not been really tested.

In this context, very recently, some of the
present authors have reported the first theoretical
investigation of the size-asymmetric
spherical electrical double layer (SEDL) \cite{Gue01}, where it was
found that, contrary to the accepted common opinion, for large macrosphere's
charge densities {\it the counterions
do not dominate}. As a matter of fact, coions are so important that their
size can induce drastic correlations that bring forth considerable
changes in the EDL's potential-charge relationship and the surge of
the charge reversal phenomenon 
in monovalent salts. Remarkably, in the
same Ref. \cite{Gue01}, the correctness of
the novel hypernetted chain/mean spherical approximation
(HNC/MSA) account of the size-asymmetric SEDL was already verified,
{\it at the level of the radial distribution functions}  
(RDFs), after comparing favorably a few 
Monte Carlo and molecular-dynamics simulations of the ionic density profiles
with the corresponding HNC/MSA integral equation results. Nevertheless, even
if this positive checking of the HNC/MSA RDFs foresees that other ensuing
theoretical predictions (e.g. the non-dominance of the counterions and  
the anomalous behavior of the electrostatic potential at the outer 
Helmholtz plane) could be true, it would be beneficial to have a specific
and more exhaustive delving of these new features by means of refined
computer ``experiments'' and/or alternative theories (i.e. integral equations,
density functionals or mean electrostatic potential schemes). Precisely, the
primary objective of this communication is to extend the
research of the size-asymmetric SEDL of Ref. \cite{Gue01} by
providing fresh and comprehensive simulational and theoretical information
that corroborates 
the enhancement of the neutralization and the screening previously found by the theory 
and, principally, the non-dominance of counterions at high colloidal charges.

The structure of this paper is as follows: the molecular model of the SEDL, theories 
and Monte Carlo simulations are described in Sec. II. 
In Sec. III the partial effects of the electrolytic size asymmetry 
in the nonlinear Poisson-Boltzmann scheme via, the URMGC approach are discussed, in order 
to establish a stand point to compare and discuss the role of the ionic size and size asymmetry 
when these features are included {\it consistently} in the MC simulations and in the integral 
equations theory, in 
the HNC/MSA and HNC/HNC approximations. To end, a summary of the main findings and 
some concluding remarks are given in Sec. IV. 

\section{MODEL AND THEORY}
\subsection{Molecular model}

The main results of this paper are based on the following representation 
of the spherical electrical double layer (SEDL), which is 
constituted by a rigid, charged spherical colloid 
of diameter $D$ and surface charge density $\sigma_0$, surrounded by a
continuum solvent of dielectric constant $\epsilon$. The macroion is
in contact with two ionic species, which are treated as hard spheres
of {\it diameters} $R_i$ ($i=1,2$) with embedded point charges of 
valences $z_i$ at their centers. 
Without loss of generality, we consider that $R_2 \geq R_1$. 
The interaction potential between the macroion, $M$,  and an
ion of type $i$ is then given by

\begin{equation}
\label{umi}
U_{Mi}(r)   = \left\{ \begin{array}{cc}
\infty, & r <  \frac{D+R_i}{2},  \\*[.2cm]
\frac{z_i e \left(\frac{D}{2}\right)^2 \sigma_0}{\epsilon_0 \epsilon r}, & 
r \geq \frac{D+R_i}{2},
\end{array}
\right. 
\end{equation}

\noindent where $e$ is the protonic charge. In turn, the interaction potential 
between two ions of species $i$ and $j$ is given by 

\begin{equation}
\label{uij}
U_{ij}(r)   = \left\{ \begin{array}{cc}
\infty, & r < \frac{R_i+R_j}{2},  \\*[.2cm]
\frac{z_i z_j e^2}{ 4 \pi \epsilon_0 \epsilon r}, & 
r \geq \frac{R_i+R_j}{2}.
\end{array}
\right. 
\end{equation}

In the classic literature, the Stern layer or, more properly, the Helmholtz {\it surface} 
is the geometrical place corresponding to the closest approach
distance between the electrolyte ions and the colloid. If 
we consider an electrolyte formed by a pair of ionic species of 
{\it unequal} size, the inner Helmholtz plane (IHP) is determined by $(D+R_1)/2$, 
i.e. by the closest approach distance of the smallest component to the surface, 
whereas the outer Helmholtz plane (OHP) is established by $(D+R_2)/2$, 
which corresponds to the distance of closest approach 
for the largest species. In the limit of identical sizes the IHP and OHP 
coincide and the usual definition of the Helmholtz plane is recovered.

\subsection{The HNC/HNC and HNC/MSA integral equations for the PM-SEDL}

The structural properties of the electrical double layer can be 
obtained from the Ornstein-Zernike equation for a multicomponent mixture of {\it S} 
species, which is: 

\begin{equation}
\label{ec1}
h_{ij}(r_{12}) = c_{ij}(r_{12}) +
\sum_{l=1}^S \rho_l \int 
h_{il}( r_{13} )c_{lj}( r_{32} ) 
dV.
\end{equation}

The set of equations  (\ref{ec1}) requires a second relation, or closure, for the functions  
$h_{ij}(r)$ and  $c_{ij}(r)$. For charged systems, the HNC and MSA closures are 
widely used. These relations are given by 

\begin{equation}
\label{ec2}
c_{ij}(r_{12}) =  
-\beta U_{ij}(r_{12}) + h_{ij}(r_{12})
- ln ( h_{ij}(r_{12}) + 1 ),  
\end{equation}

\noindent for HNC, and 

\begin{equation}
\label{ec3}
c_{ij}(r_{12}) =  -\beta U_{ij}(r_{12}), 
\end{equation}

\noindent for MSA, with

\begin{center}
$i,j=1,2 \dots S.$ 
\end{center}

These expressions are complemented by the exact condition $h_{ij}=-1$ when 
$r_{ij}<R_{ij}$, such as $R_{ij}=(R_i+R_j)/2$. 

Let us consider that the species {\it S} 
corresponds to macroions at infinite dilution in a binary 
electrolyte. Then Eqs. (\ref{ec1}) for species 
{$S \equiv M$} and $j$ can be written as:

\begin{equation}
h_{Mj}(r_{12}) = c_{Mj}(r_{12}) +
\sum_{l=1}^2 \rho_l \int 
h_{Ml}( r_{13} )c_{lj}( r_{32} ) 
dV.
\label{ec4}
\end{equation}

\begin{center}
$j=1,2.$ 
\end{center}

When Eq. (\ref{ec2}) is used in (\ref{ec4}) for both $c_{Mj}(r)$ and $c_{ij}(r)$ 
the HNC/HNC integral equation is obtained for the SEDL. 
Besides, if Eq. (\ref{ec2}) is employed in Eq. (\ref{ec4}) only for $c_{Mj}(r)$,  
and the $c_{ij}(r)$ are approximated by the corresponding MSA bulk expressions 
(i.e. Eq. (\ref{ec3}) is inserted in Eq. (\ref{ec1}) for the two electrolytic species), 
the HNC/MSA integral equation is established. 
The details of these integral equations formalisms can be consulted 
elsewhere \cite{Pa01,Go02,Gue01} and will not be repeated here. However, 
it is important to mention that both schemes 
satisfy the global electroneutrality condition.

As an especial case of HNC/MSA, if $R_1=R_2=0$ in Eq. (\ref{uij}), and $R_1 \neq 0$ and 
$R_2 \neq 0$ in Eq. (\ref{umi}), this equation reduces to the integral equation 
version of the nonlinear 
Poisson-Boltzmann (PB) equation \cite{Gue01}:

\begin{equation}
\nabla^2  \psi = - \frac{1}{\epsilon_0 \epsilon} \sum_{i=1}^2 z_i e \rho_i exp(-z_i e \psi/ k_B T).  
\label{pb_ec}
\end{equation}

\noindent When $R_1$ and $R_2$ are equal in Eq. (\ref{umi}), Eq. (\ref{pb_ec}) 
is the so-called the Modified Gouy Chapman equation (MGC). 
On the contrary, if $R_1$ and $R_2$ are different in Eq. (\ref{umi}), 
the nonlinear PB equation corresponds to the unequal-radius MGC (URMGC) 
equation \cite{Va01}. Notice that in these quasi point-like ions theories the ionic size 
is only taken into account 
partially at the level of a closest approach distance between the macroion and the ions. 
Contrastingly, in the MC simulations and in the HNC/MSA and HNC/HNC integral equations 
the ionic size and size asymmetry are incorporated consistently, 
via the primitive model. 

To better understand and characterize the behavior of the SEDL in presence of the 
electrolytic size asymmetry, 
from the radial distribution functions, $g_{ij}(r)$,  
it is possible to calculate two important quantities, namely, the integrated charge (IC)

\begin{equation}
P(r) = z_M + \sum_{i=1}^2 \int_0^r z_i \rho_i g_i(r) 4 \pi r^2 dr, 
\label{ic}
\end{equation}

\noindent and the mean electrostatic potential (MEP)

\begin{equation}
\psi(r) =   \frac{e}{4 \pi \epsilon_0 \epsilon} \int_r^\infty  \frac{P(t)}{t^2} dt.
\label{pot2}
\end{equation}

\noindent When the MEP is evaluated at $r=(D+R_1)/2$
Eq. (\ref{pot2}) corresponds to the MEP at the IHP, which we denote as $\psi_{IHP}$. 
On the other side, if Eq. (\ref{pot2}) is calculated at $r=(D+R_2)/2$ the MEP at the OHP, 
$\psi_{OHP}$, is obtained. 

With respect to the IC, this quantity is a measure of the 
total net charge inside a sphere of radius $r$ 
centered in the macroion.
Then, if $D/2 \leq r \leq (D+R_1)/2 $ the IC is equal to $z_M$,
whereas for {$r \rightarrow \infty$}  this quantity goes to zero 
because of the electroneutrality condition. 
Furthermore, the IC has also the property 
of indicating charge reversal when $P(r)z_M < 0$. 

\subsection{Numerical simulations}

Monte Carlo simulations of the SEDL were performed considering a 
cubic box with a macroion fixed at the center
under periodic boundaries. Due to 
the electroneutrality condition the following relation was satisfied:

\begin{equation}
N_-z_- + N_+z_+ + z_M = 0,
\end{equation}

\noindent where $N_-$ and $z_-$ are the number of ions and the valence of 
the negative species, respectively, 
$N_+$ and $z_+$ are the number of ions and the valence corresponding to the positive 
species, and $z_M$ is the valence of the macroion, which is related to the surface 
charge density as $\sigma_0=z_Me/ \pi D^2$. 

In order to take into account the long range nature of the 
coulombic potential, the Ewald sums scheme was adopted, using conducting 
boundary conditions \cite{Al01,Fr01}. The damping constant $\alpha$ was set 
to $\alpha=5/L$ and the $k$-vectors employed  to compute the reciprocal space 
contribution to the energy satisfied the condition $k \leq 5$. 
The length $L$ of the simulation box was assigned considering a total number of ions 
$N_t = N_- + N_+ \approx 1000$. After ${N_t}$ attempts to move an arbitrary ion a 
Monte Carlo cycle is counted. 
The thermalization process consisted of $2 \times 10^4$ MC cycles, and from 
$2 \times 10^6$ (for high $z_M$ values) to $6 \times 10^6$ (for low $z_M$ values) MC 
cycles were completed in order to calculate the canonical average. The 
quality of the simulation was tested calculating the IC, which in 
a region far from the macroion and near of the borders 
of the simulation box vanished in all cases, as expected. 

\begin{figure}[htbp]
\begin{center}
\begin{minipage}{\linewidth}
{\includegraphics[angle=0.0, width=\linewidth]{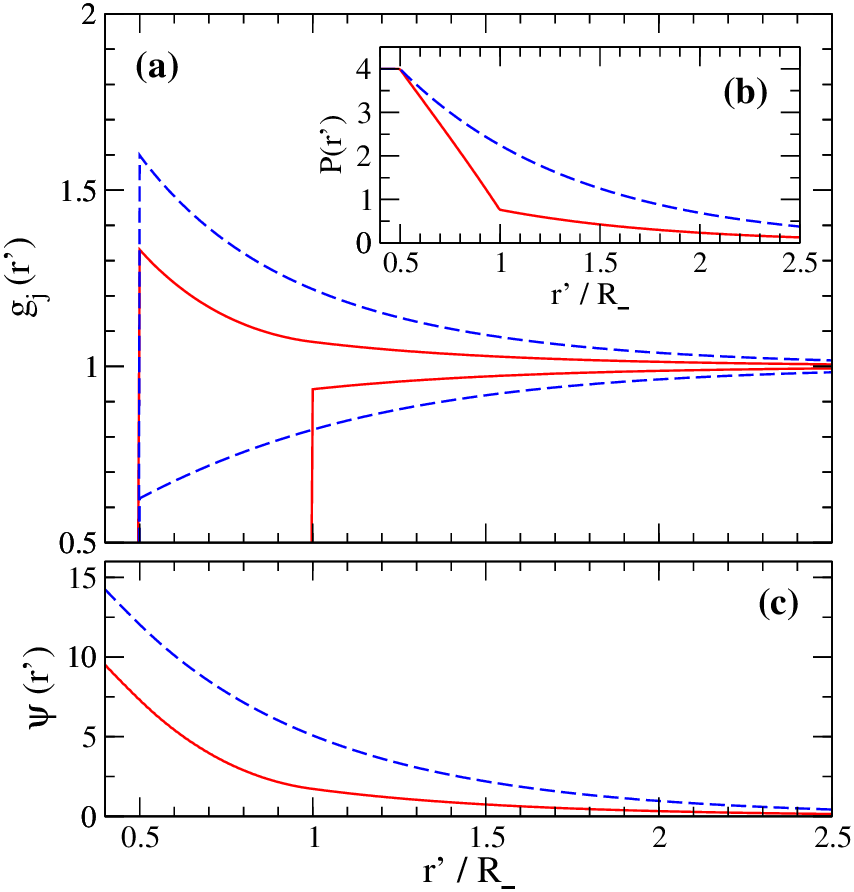}} 
\caption{(Color online:) Radial distribution functions, integrated charge and mean 
electrostatic potential 
of a size-symmetric and size-asymmetric 1:1 salt around a charged macroion of valence 
$z_M=4$ ($\sigma_0=0.05$~$C/m^2$) and diameter $D=20$~\AA~ in the PB approach.  
The solid and dashed lines correspond to URMGC and MGC equations, respectively. 
}
\label{comp1_pb}
\end{minipage}
\end{center}
\end{figure}

\begin{figure}[htbp]
\begin{center}
\begin{minipage}{\linewidth}
{\includegraphics[angle=0.0, width=\linewidth]{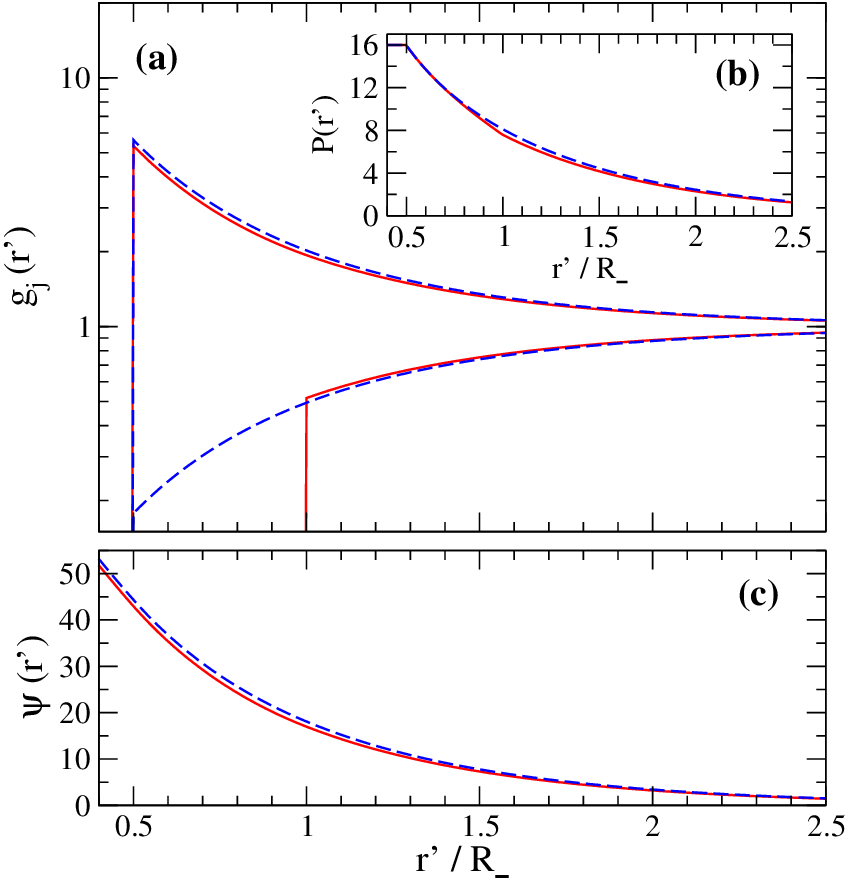}}
\caption{(Color online:) The same as in Fig. \ref{comp1_pb} but for 
$z_M=16$ ($\sigma_0=0.2$~$C/m^2$). 
}
\label{comp2_pb}
\end{minipage}
\end{center}
\end{figure}

\begin{figure}[htbp]
\begin{center}
\begin{minipage}{\linewidth}
{\includegraphics[angle=0.0, width=\linewidth]{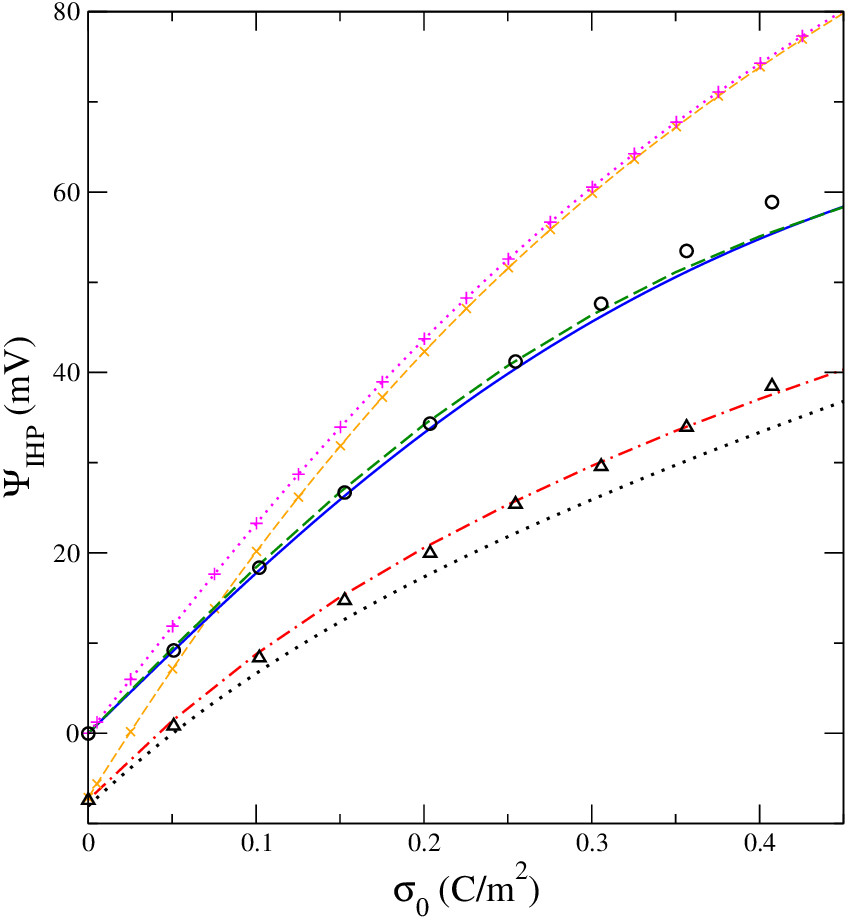}}
\caption{(Color online:) Mean electrostatic potential at the IHP as function of 
the surface charge density $\sigma _{0}$, for a 1:1, 1 M electrolyte around 
a macroion of diameter $D=20$~\AA.
The simulations results were calculated for $z_M=0,4,8,12,16,20,24,28$ and $32$.
The diameters of the ionic species in the PM are $R_{-}=4.25$~\AA\ and $R_{+}=8.5$~\AA. 
In the RPM, the ionic diameters are equal to the counterions 
in the PM, i.e. $4.25$~\AA\ . The maximum approach distances $d_{ij}$ 
for ion-ion and macroion-ion interactions for theory and simulation 
are given by Eqs. (\ref{max_col_ion_sim}) y  (\ref{max_ion_ion_sim}).  
The triangles and the circles correspond to Monte Carlo simulation results in the 
PM, MC$_{PM}$, and in the RPM, MC$_{RPM}$, respectively. 
The dotted and solid lines correspond to HNC/MSA$_{PM}$ and HNC/MSA$_{RPM}$, and 
the dot-dashed and dashed lines are associated to HNC/HNC$_{PM}$ and HNC/HNC$_{RPM}$, 
respectively. 
The dashed line with multiplication symbols denotes URMGC and the dotted line 
with plus symbols is for MGC. 
}
\label{z_c2_ihp}
\end{minipage}
\end{center}
\end{figure}

\begin{figure}[htbp]
\begin{center}
\begin{minipage}{\linewidth}
{\includegraphics[angle=0.0, width=\linewidth]{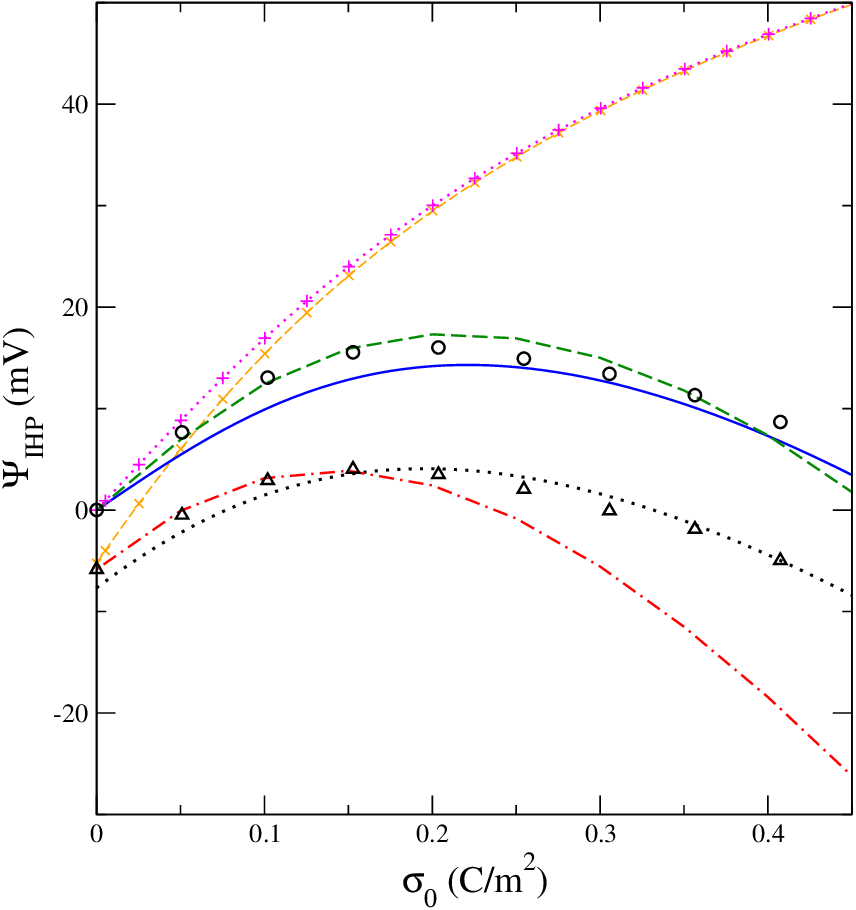}}
\caption{ 
(Color online:) The same as in Fig. \ref{z_c2_ihp} but for a 2:2 electrolyte 
at 0.5 M concentration. 
}
\label{z_c6_ihp}
\end{minipage}
\end{center}
\end{figure}

\begin{figure}[htbp]
\begin{center}
\begin{minipage}{\linewidth}
{\includegraphics[angle=0.0, width=\linewidth]{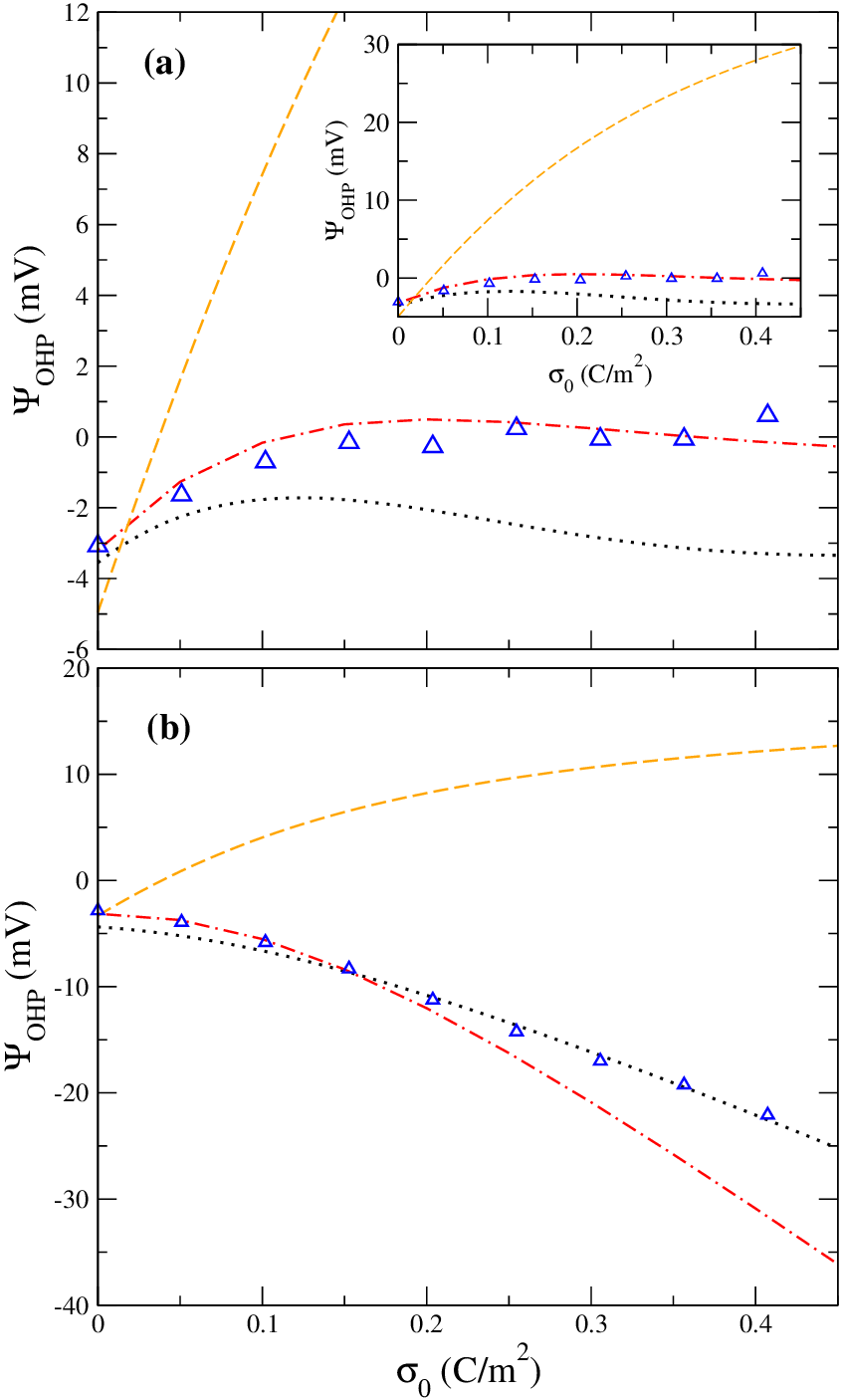}}
\caption{
(Color online:) Mean electrostatic potential at the OHP as function of 
the surface charge density $\sigma _{0}$, around a 
macroion of diameter $D=20$~\AA.
The simulation results were calculated for $z_M=0,4,8,12,16,20,24,28$ and $32$.
The diameters of the ionic species in the PM are $R_{-}=4.25$~\AA\ and $R_{+}=8.5$~\AA. 
The maximum approach distances $d_{ij}$ for ion-ion and macroion-ion 
interactions for theory and simulation 
are given by Eqs. (\ref{max_col_ion_sim}) y  (\ref{max_ion_ion_sim}).  
The triangles correspond to Monte Carlo simulations. The dotted, dot-dashed 
and dashed are associated to HNC/MSA$_{PM}$ and HNC/HNC$_{PM}$, 
and URMGC, respectively. In Fig. \ref{z_c2_c6_ohp}a the electrolyte is 1:1, 1 M, 
whereas in Fig. \ref{z_c2_c6_ohp}b the salt is 2:2, 0.5 M.} 
\label{z_c2_c6_ohp}
\end{minipage}
\end{center}
\end{figure}

\begin{figure}[htbp]
\begin{center}
\begin{minipage}{\linewidth}
{\includegraphics[angle=0.0, width=\linewidth]{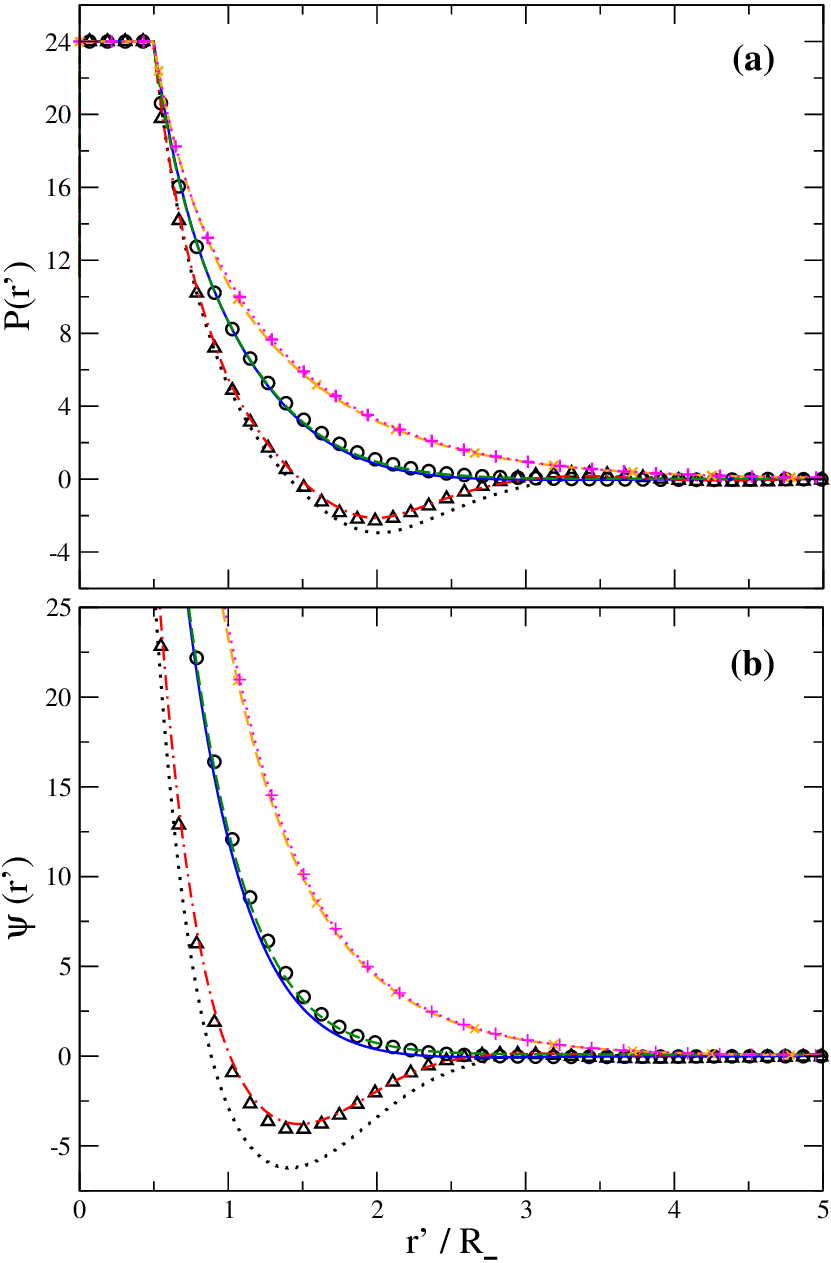}}
\caption{
(Color online:) Integrated charge and mean electrostatic potential 
as function of the distance for a 1:1, 1 M electrolyte 
near a macroion of diameter $D=20$ \AA\ and $z_M= 24$.  
The diameters of the ionic species in the PM are $R_{-}=4.25$~\AA\ and $R_{+}=8.5$~\AA. 
In the RPM, the ionic diameters are equal to the counterions 
in the PM, i.e. $4.25$~\AA. The maximum approach distances $d_{ij}$ for 
ion-ion and macroion-ion interactions for theory and simulation 
are given by Eqs. (\ref{max_col_ion_sim}) y  (\ref{max_ion_ion_sim}). 
The triangles and the circles correspond to 
MC$_{PM}$ and MC$_{RPM}$, respectively.
The dotted and solid lines correspond to HNC/MSA$_{PM}$ y HNC/MSA$_{RPM}$, and 
the dot-dashed and dashed lines are associated to HNC/HNC$_{PM}$ and HNC/HNC$_{RPM}$, 
respectively. 
The dashed line with multiplication symbols denotes URMGC and the dotted line 
with plus symbols is for MGC. The distance $r$' is measured from the macroion's surface.}
\label{qr_vr_c2}
\end{minipage}
\end{center}
\end{figure}

\begin{figure}[htbp]
\begin{center}
\begin{minipage}{\linewidth}
{\includegraphics[angle=0.0, width=\linewidth]{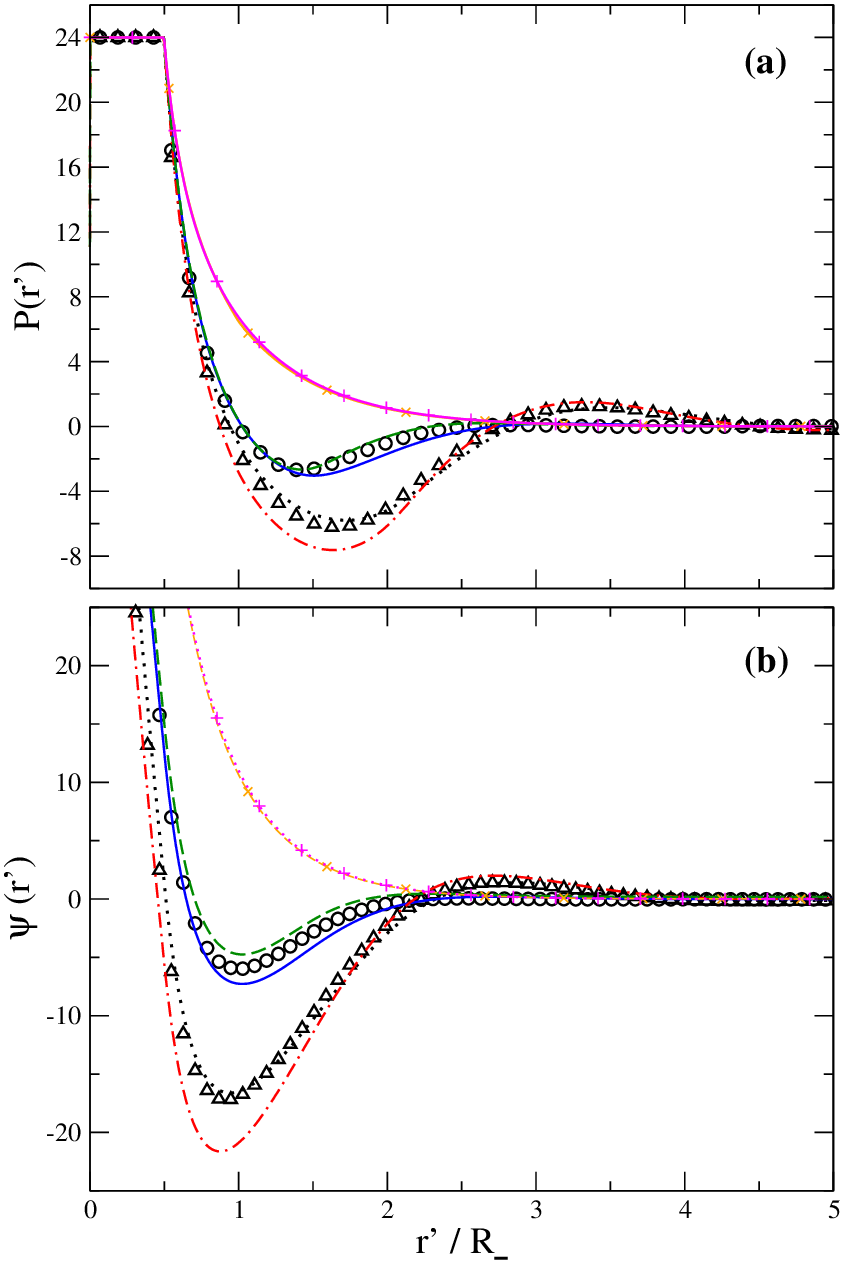}}
\caption{
(Color online:) The same as in Fig. \ref{qr_vr_c2} but for a 2:2, 0.5 M salt.
}
\label{qr_vr_c6}
\end{minipage}
\end{center}
\end{figure}

\section{RESULTS AND DISCUSSION}

\subsection{The role of the ionic size asymmetry in the PB scheme}

Physically, in the MGC and URMGC equations the ionic size is considered only partially because 
the electrolytic ions are allowed to be close to the macroion until a closest approach distance 
for each species, but the ions interact among them as charged points. Thus, 
the MGC and URMGC equations correspond, to the simplest manner in which the effects 
of the ionic size and 
the size asymmetry can be studied in the EDL, as it was done by Valleau et al. 
in the 1980s for the planar instance \cite{Va01}. Notwithstanding, 
although URMGC represented a step beyond MGC, when the ionic size and in 
particular the ionic size asymmetry is fully taken into account (as occurs in 
the MC simulations and in the HNC/MSA and HNC/HNC integral equations) new features 
absent in PB picture emerge. Therefore, in order to later compare and discuss 
the consequences of a complete consideration of ionic size asymmetry in the SEDL, 
we will begin with a review of size asymmetry in the URMGC approach, 
where the excluded volume effects are embodied partially 
only in the colloid-ion interactions. Thus, let us first study  
a spherical macroion of diameter and surface charge density $D=20$~ \AA~ and
$\sigma_0=z_Me/ \pi D^2$, respectively , surrounded by a 1:1, 1 M electrolyte, 
in a continuum solvent of dielectric constant $\epsilon=78.5$ at a temperature 
$T=298$ $K$. 
In the size-symmetric case (i.e. for the MGC theory), 
the maximum approach distance for both species 
is $4.25/2$~\AA~, whereas in the 
size-asymmetric (i.e. for the URMGC theory) is $4.25/2$~\AA~ for anions and  
$8.5/2$~\AA~ for cations. 
Since we will consider only $\sigma_0>0$ values,  
in both instances the counterions have the same properties, 
being the size of the coions the unique difference between the MGC and URMGC systems. 
Therefore, in Fig. \ref{comp1_pb} we compare the RDFs, ICs and MEPs curves associated to the 
size-symmetric and to the size-asymmetric 
cases, when the valence of the macroion is $z_M=4$ ($\sigma_0=0.05 C/m^2$). 
At the level of the RDFs, in Fig. \ref{comp1_pb}a it is observed that the  
MGC profiles for the size-symmetric case enclose the RDFs of the size-asymmetric  
electrolyte described by URMGC. Besides, from Figs. \ref{comp1_pb}b and 
\ref{comp1_pb}c it is seen 
that the region not allowed for big cations but accessible for the small anions in the 
URMGC theory contributes significantly to the increase of the neutralization 
and the screening of the spherical EDL when contrasted with the MGC theory 
(notice that $P(r)_{URMGC} \leq P(r)_{MGC}$ and $\psi(r)_{URMGC} \leq \psi(r)_{MGC}$
for all the $r$ plotted). 
These differences in the RDFs, $P(r)$, and $\psi(r)$  
are expected to augment if $z_M$ decreases, 
with the largest dissimilarities occurring precisely at the point of zero charge (PZC).  
Additionally, it is foreseen a MEP equal to zero at the closest approach distance of anions
in the MGC results and an electrostatic potential different from zero for URMGC at the same distance.
This happens for a $z:z$ salt when $z_M=0$ because, under these conditions,  
the RDFs for MGC must coincide for symmetry reasons, while for URMGC the ionic density 
profiles must be completely different in order to fulfill the electroneutrality 
condition (i.e. 
as the macroion is uncharged, only the accumulation 
of the large ionic species after the OHP 
can compensate the adsorbed charge of the small ions in 
between the Helmholtz planes). On the contrary, if 
$z_M$ increases and there are no crossings between the RDFs of counterions and coions 
for MGC, and the same behavior is displayed by URMGC, the MGC curves 
are the limit of URMGC profiles when $z_M \to \infty$ 
due again to the electroneutrality condition. 
Such phenomenon has been already discussed in \cite{Gue01}. 
Thus, even if, strictly, the MGC and URMGC profiles should not be the same for high 
$z_M$ values (because there is always a region where the small 
coions can exist in URMGC, see Fig.\ref{comp2_pb}a), 
the structural properties of the EDL, as the integrated charge 
and the mean electrostatic potential,  
are indeed very similar as can be observed in Figs. \ref{comp2_pb}b and \ref{comp2_pb}c. 
Consequently, far from the PZC the properties of the spherical EDL are expected to be practically 
the same for URMGC and MGC 
 because the coion's contribution is negligible and 
the counterions are the same in both cases. This last merging between URMGC and MGC 
is precisely the so-called 
dominance of counterions in the spherical EDL and will be of decisive importance in 
the corresponding potential-charge relationship, as it will be shown later. 

\subsection{The role of the ionic size asymmetry in the primitive model: 
MC simulations and theoretical results}

As discussed, one of the main differences between 
size-symmetric and size-asymmetric EDLs salt in the PB viewpoint   
is the increment in the neutralization or screening predicted by URMGC,  
with respect to MGC, near the PZC. In addition, for high surface charge densities 
the properties of the SEDL are expected to be practically the same,  
such as the dominance of counterions arises in a quasi-point like description (i.e. MGC and URMGC). 

Now, in this section we will show comprehensive MC data and integral equations (IE) results 
in which the ionic size asymmetry is taken into account {\it consistently} in the Eqs. 
(\ref{umi}) and (\ref{uij}) (and not only at the macroion-ion level as it was done 
in the nonlinear PB equation of Sec. III A) in order to display the effects in the 
SEDLs due to a more realistic treatment of size-symmetric and size-asymmetric salts. 

In all the following simulations and theoretical calculations 
we considered a macroion of diameter $D=20$ \AA~ and $\sigma_0 \ge 0$, 
immersed in a continuum solvent of dielectric constant $\epsilon=78.5$ at a 
temperature $T=298$ $K$, 
in presence of a binary electrolyte. In the primitive model (PM) the {\it diameter} of the 
counterions is $R_-=4.25$ \AA~ and the 
diameter of coions is the double, i.e. $R_+=8.5$ \AA. For the restricted 
primitive model, the size of both species 
is the same of the counterions in the PM, i.e. $4.25$ \AA. Thus, the maximum approach distances 
in the PM and the RPM correspond to those of the electrolyte for URMGC and MCG, respectively. 
This information is summarized in Eqs. (\ref{max_col_ion_sim}) and 
(\ref{max_ion_ion_sim}). 
Notice that the diameter of the macroion and the ionic size asymmetry correspond to  
values that emphasize the spherical geometry of the EDL and that have been  
typically used in previous works \cite{Gr02,Wu01,Te01,Gue01}. 

\begin{table*}

\begin{equation}
\label{max_col_ion_sim}
d_{i}=\left\{
\begin{array}{ll}
d_{-}=\frac{D+R_{-}}{2}~ \textnormal{and}  ~d_{+}=\frac{D+R_{+}}{2}\textnormal{,} &
 \textnormal{for MC}_{PM} \textnormal{, } \textnormal{HNC/MSA}_{PM} \\
& \textnormal{HNC/HNC}_{PM} \textnormal{ and URMGC;} \\
d_{-}=d_{+}=\frac{D+R_{-}}{2}\textnormal{,} & \textnormal{for MC}_{RPM} \textnormal{, }
\textnormal{HNC/MSA}_{RPM} \\
&\textnormal{HNC/HNC}_{RPM} \textnormal{and MGC;}
\end{array}
\right.
\end{equation}

\begin{equation}
\label{max_ion_ion_sim}
d_{ij}=\left\{
\begin{array}{ll}
d_{-\,-}=R_{-}~ \textnormal{and}  ~d_{+\,+}=R_{+}\textnormal{,} 
& \textnormal{for MC}_{PM} \textnormal{ , }
\textnormal{HNC/MSA}_{PM} \textnormal{ , } \textnormal{HNC/HNC}_{PM}
\textnormal{;} \\
d_{-\,+}=d_{+\,-}=\frac{R_{-}+R_{+}}{2}\textnormal{,} 
& \textnormal{for MC}_{PM} \textnormal{ , }
\textnormal{HNC/MSA}_{PM} \textnormal{ , } \textnormal{HNC/HNC}_{PM}
\textnormal{;} \\
d_{-\,-}=d_{+\,+}=d_{-\,+}=d_{+\,-}=R_{-}\textnormal{,} 
& \textnormal{for MC}_{RPM} \textnormal{ , }
\textnormal{HNC/MSA}_{RPM} \textnormal{ and } \textnormal{HNC/HNC}_{RPM}
\textnormal{;} \\
d_{-\,-}=d_{+\,+}=d_{-\,+}=d_{+\,-}=0\textnormal{,} 
& \textnormal{for URMGC and MGC.}
\end{array}
\right.
\end{equation}

\end{table*}

Very importantly, the mean electrostatic potential at some distance from the 
surface of the macroion is frequently associated 
with the so-called electrokinetic potential at the shear plane (or the 
zeta potential, $\zeta$) \cite{Hu02}. Such quantity is very relevant in colloidal 
studies since it is experimentally measurable 
and allows to characterize and summarize the behavior of the SEDL, as functions 
of the colloidal charge, in a single potential-charge plot. 
Furthermore, the zeta potential is used often in physical-chemistry 
to characterize the macroscopic properties and the stability of charged  
colloidal dispersions \cite{Fe01,Hi01}. 
Given that, in the past, the $\psi_{IHP}$ or the $\psi_{OHP}$ have identified with $\zeta$, 
we start by showing the $\psi$-$\sigma_0$ curves 
at the IHP and at the OHP for our PM systems. 
In Fig. \ref{z_c2_ihp}, MC simulations of the mean electrostatic potential at the IHP for a 
1 M, 1:1 electrolyte in the PM and the RPM are shown. 
The first notable feature there is the {\it merging} of the MGC and URMGC curves for high 
$\sigma_0$ values. Precisely, this asymptotic conduct illustrates 
the dominance of counterions at the level of $\psi$-$\sigma_0$. Besides, in this figure 
it is seen that the maximum difference between the MEPs corresponding to MGC and URMGC 
happens precisely at the PZC, as it had been pointed out in Sec. III A.   
In strong contrast, the most evident characteristic displayed by the simulational data  
of $\psi_{IHP}$ 
for the size-symmetric (RPM) and size-asymmetric (PM) instances is that these curves   
{\it do not converge to the same one 
when $\sigma_0$ augments. This confirms that the 
counterions do not always dominate the EDL far from the PZC, or, in 
other words, exemplify the importance of the size of the coions at high colloidal 
charges, as it was theoretically predicted in Ref. \cite{Gue01}}. 
Additionally, the simulational data shows a potential different from zero 
at the PZC for the size-asymmetric electrolyte. 
Clearly, such behavior is due to the fact that the small negative ions are 
allowed to be closer to the surface than the big positive ions, i.e. for $\sigma_0=0$ the 
negative sign of the MEP at the IHP results from the size asymmetry of the 1:1 
electrolyte. 
Interestingly, analogous results 
had been theoretically predicted for the RPM planar EDL 
of charge asymmetric species \cite{Bo03,Bh02}. However, 
these data have not been confirmed simulationally. 
With respect to the performance of the integral equation theories  
in the RPM, it is remarkable that both the HNC/HNC and HNC/MSA theories agree with the simulation data, 
although, for the PM case, HNC/HNC follows closely the MC curves than HNC/MSA.  

In addition, in Fig. \ref{z_c2_ihp} the MC simulations and integral equations 
(IE) predict a larger value of 
$\sigma_0$ for which the negative sign of the MEP at the IHP 
changes to positive than that corresponding to URMGC. This exemplify again the 
importance of the entropic contributions in the adsorption of counterions 
between the Helmholtz planes when the ionic size correlations are 
considered consistently (as occurs in the MC simulations), in contrast 
with its partial inclusion, when only different approach distances 
in the macroion-ion interaction are considered, which neglects ionic excluded volume 
effects outside of the OHP (as in the URMGC theory). Another phenomenon 
observed in Fig. \ref{z_c2_ihp} is that $\psi(\sigma_0)_{PM} < \psi(\sigma_0)_{RPM}$  
at the IHP for all $\sigma_0$ values plotted. Near the point of zero charge this 
is explained, in the PM, in terms of the adsorption of negative counterions 
that are not neutralized by the big coions, as it happens in the RPM  
case. When $\sigma_0$ augments, the positive bare charge overcomes 
the contribution of the negative counterions near the macroion's surface and the MEP's sign 
changes from negative to positive as it was mentioned above. Nevertheless, 
for all the $\sigma_0$ values displayed 
the counterions in the PM provide an extra screening which is not present in the RPM, 
which leads to the $\psi(\sigma_0)_{PM} < \psi(\sigma_0)_{RPM}$ condition. 
Moreover, in the monovalent case of the PM this extra screening can be related not only with a higher 
adsorption of negative counterions with respect to the RPM but
also with the presence of charge reversal, that is absent
in the RPM. This behavior will be clarified later when the corresponding $P(r)$ profiles be 
presented.

With the purpose of performing a more stringent test for the theories, 
in Fig. \ref{z_c6_ihp} we present MC simulations of the MEP at the IHP 
for a 0.5 M, 2:2 electrolyte in the PM and the RPM. In this instance, 
the features already observed in the simulations of monovalent 
ions are accentuated. In particular, the most important finding is the corroboration of  
the non-dominance of counterions for divalent ions. 
In addition, in the size-asymmetric case 
a very strong adsorption of negative 
counterions in the PM is also observed. The importance of excluded volume effects  
is evinced by noticing that near the PZC the interval 
of $\sigma_0$ for which the $\psi_{IHP}$ is negative is larger for 
MC simulations and IE theories than for URMGC. 
Furthermore, when $\sigma_0$ increases after $\psi(\sigma_0)_{PM} > 0$,  
the $\psi_{IHP}$ reaches a maximum and for still larger $\sigma_0$ 
values $\psi(\sigma_0)_{PM} < 0$ again. The appearance of a maximum in the 
$\psi(\sigma_0)$ plot for the RPM is related to presence 
of charge reversal. 
Besides, the early MEP's change of sign in the PM after the maximum 
displayed by MC simulations suggests an extra adsorption of counterions with 
respect to the RPM case, 
i.e. it is expected an accentuated charge reversal that screens more strongly the positive 
bare charge of the macroion for high $\sigma_0$ values. 
This behavior will be clearly exhibited in the corresponding $P(r)$ 
profiles later. On the other hand, the simulational confirmation of the reentrance 
in the sign of the MEP for divalent ions in the PM, i.e. 
the double change of sign of $\psi_{IHP}$, 
suggests the possibility of observing 
a corresponding reentrance in the experimental electrophoretic mobility ($\mu$), 
if the Smolouchowski equation 
($\mu = \epsilon \zeta / \eta$) is valid, 
as it had been theoretically foreseen by HNC/MSA \cite{Gue01} for a larger macroion. 
This means that the ionic size asymmetry could then cause a reversed mobility in the 
motion of a macroion 
in an electrophoresis experiment near the PZC, changing to the ``correct'' direction 
when its surface charge density augments, but inverting again its movement at high 
$\sigma_0$ values. 
Also, notice that for the size-asymmetric instance HNC/HNC shows a better 
agreement with the simulation data 
than HNC/MSA for low colloidal charges (when $\sigma_0 \leq 0.16$~$C/m^2$ approximately). 
Contrastingly, for high surface charge densities ($\sigma_0 < 0.16$~$C/m^2$) the opposite 
behavior is observed, i.e. 
a substantial deviation from the MC data is displayed by HNC/HNC, 
which contrasts with the good accordance shown by HNC/MSA. This conduct is remarkable since, 
for high colloidal charges and high entropic-electrostatic ionic correlations, 
in the divalent case HNC/MSA is better than HNC/HNC in opposition to the univalent instance.

In Fig. \ref{z_c2_c6_ohp} the simulational results of the MEP at the OHP, $\psi_{OHP}$, 
for the PM monovalent and divalent salts are plotted. 
As we said before, the importance of the study of this curves arises from the fact 
that the $\psi_{OHP}$ could be identified with the zeta potential ($\zeta$). 
In Fig. \ref{z_c2_c6_ohp}a, which corresponds to the 1:1, 1 M electrolyte, 
a non-monotonic behavior of the MEP 
as a function of $\sigma_0$ is observed. This conduct 
is reproduced correctly by IE theories,  with HNC/HNC showing 
a better agreement than HNC/MSA. Contrastingly, URMGC presents a monotonic behavior,  
which completely differs from the simulation data, and predicts  
large values of the electrostatic potential for high $\sigma_0$ as it can be 
seen in the inset. The simulational results of $\psi_{OHP}$ 
for the 2:2, 0.5 M electrolyte are portrayed in Fig. \ref{z_c2_c6_ohp}b. 
Here, in the MC simulations it is observed that for any value of $\sigma_0$ 
the $\psi_{OHP}$ is negative 
and decreases monotonically as a function of $\sigma_0$. A similar curvature 
has been theoretically predicted in the RPM spherical EDL by 
one of the present authors for 
a macroion of diameter $D=80$~\AA~ immersed in a 2:2, 0.5 M electrolyte of 
ionic diameter equal to 7~\AA~ \cite{Go03}, at approximately
the same volume fraction used here in our PM.
Thus, the simulational results presented in this work corroborate that such anomalous 
curvatures in the MEP,    
at high ionic volume fractions, are a real feature in the primitive model.  
Even if this phenomenon is interesting just from the theoretical point of view \cite{Blu01,Blu02,Att01},  
could also be relevant in the description of non-intuitive attributes of double layer systems 
such as the occurrence of negative differential capacitances \cite{Att01,To01,Par01,Par02,Ki01,Par03}.  
Once again, the whole behavior displayed by the simulations  
is well captured by the the IE theories, although now HNC/MSA 
is closer to simulations than HNC/HNC. Contrastingly, URMGC exhibits a monotonic 
behavior in which the MEP increases 
as a function of $\sigma_0$, as occurred in the 1:1 instance.

As was noticed for the $\psi(\sigma_0)$ relationship, one of the consequences of 
the ionic size asymmetry in the PM electrical double layer, is the enhancement of the 
neutralization and screening at high surface charges.
To illustrate this in terms of the ionic charge adsorption, in Fig. \ref{qr_vr_c2}a 
we have plotted the integrated charge profiles for the 1:1, 
1 M electrolyte in the PM and RPM, when the macroion's valence is $z_M=24$. 
Here, it is clearly seen that the ionic size asymmetry 
not only promotes a higher adsorption of counterions, i.e.
$P(r)_{PM} \leq P(r)_{RPM}$, but also that can {\it induce the appearance of charge reversal in monovalent electrolytes}, 
showing a minimum at  
$r$'$/R_- \approx 2$. Consequently, MC simulations hint that charge reversal 
can occur even in presence of monovalent salts whenever the 
high coupling conditions are present, i.e. high electrolyte concentration 
or large hydration of the electrolyte. Note that the  
IE theories reproduce very well the charge reversal behavior, 
whereas an incorrect monotonic decrease of the IC is 
shown by URMGC.

The overcompensation of the native charge in the PM is also reflected in the 
corresponding MEP curves 
as a function of the distance 
as can be verified in Fig. \ref{qr_vr_c2}b. 
In this case the monotonicity of
the $\psi(r)_{RPM}$ and the non-monotonic behavior of $\psi(r)_{PM}$ 
can be easily deduced from the corresponding $P(r)$ profiles and the Eq. (\ref{pot2}). 
In particular, note that the condition $P(r)_{PM} \leq P(r)_{RPM}$ 
implies that $\psi(r)_{PM} \leq \psi(r)_{RPM}$ for MC simulations and HNC/HNC and HNC/MSA theories, i.e. 
to a higher neutralization of the macroion's bare charge (as observed in $P(r)$) 
a higher screening in $\psi(r)$ is associated.

In the Fig. \ref{qr_vr_c6}a the charge adsorption for a 2:2, 0.5 M  
electrolyte as function of the distance in the PM and RPM is displayed 
when the valence of the macroion is $z_M=24$.  
Notice that the integrated charge simulation curve shows that in the RPM electrical double layer 
there is charge reversal, contrasting with the behavior of the 1 M monovalent case previously 
portrayed, where it was found the absence of this feature even for a higher electrolytic concentration. 
Besides, when the ionic size 
asymmetry is present in the PM the charge reversal, already observed in the RPM, is 
notably enhanced. This illustrates the fact that 
{\it the high electrostatic-entropic coupling conditions for which the charge reversal appears 
can be relaxed for multivalent salts}, i.e.,   
it is expected that for $1:z$ salts the ionic size and/or the ionic size asymmetry 
(coming from the ionic hydration for example) 
become very important even at moderate salt concentrations reachable experimentally, 
as it has been reported by several experimental works \cite{Mar01,Bes01,Bes02,Fer01,Hey01}. 
The MEP curves corresponding to the IC profiles discussed previously now   
are plotted in Fig. \ref{qr_vr_c6}b. Consistently, the MC simulations and the HNC/HNC HNC/MSA theories 
predict that near the macroion's surface the screening in the PM is higher than in the 
RPM, with the MEP presenting a non-monotonic behavior in both cases. Furthermore, 
here it is noticed that the overestimation of the screening in the HNC/HNC potential profile has its 
origin in the charge overcompensation displayed by the corresponding $P(r)$ graphed in Fig. 
\ref{qr_vr_c6}a. 

\section{CONCLUDING REMARKS}

Monte Carlo simulations of the primitive model spherical electrical 
double layer, in the presence of either monovalent 
or divalent salts, were performed and compared with data of the HNC/HNC and HNC/MSA 
integral equations and of the the MGC and URMGC quasi-punctual-ions schemes. 
One of the most simple manners in which the ionic size 
and size asymmetry can be taken into account {\it partially} in the EDL 
is via the MGC and URMGC approaches, in which the excluded volume effects are considered solely 
in the macroion-ion interactions by means of a closest approach distance. 
When size-symmetric and size-asymmetric semi-punctual EDLs (i.e. 
systems with either equal or different closest 
approach distances for the otherwise-punctual electrolytic species) with the same type of 
counterions are considered by the MGC and URMGC formalisms, respectively, 
it was exhibited here that, at low colloidal charges, one of the main effects of 
including the ionic size asymmetry 
in the URMGC theory is an enhancement of the screening and the neutralization in the EDL 
with respect to the MGC (or size-symmetric) instance. 
On the other hand, far from the point of zero charge, it was shown that the RDFs, ICs, and MEPs 
predicted by the MGC and URMGC equations displayed always a monotonic comportment, 
and that the ionic distributions almost coincided for all the values of $r$, with 
exception, maybe, of the zone comprised between 
the Helmholtz planes. However, as $\sigma \to \infty$, the diferences between 
the MGC and URMGC radial distribution functions, 
and therefore between all their concomitant structural and thermodynamic 
properties, go asymptotically to zero. This behavior is 
the so-called dominance of the counterions. 
Contrastingly, when the ionic finite size and size asymmetry 
are embodied consistenly into the macroparticle-ion and ion-ion interactions, 
as occurred in the Monte Carlo (MC) simulations 
and the HNC/HNC and HNC/MSA integral equations, several characteristics absent in the nonlinear 
Poisson-Boltzmann data (e.g. the non-montonic behavior 
of the RDFs, ICs, and MEPs and the phenomenon of charge reversal) emerge. 
Most importantly, our MC simulations of the primitive model EDL 
corroborated the fact that the ionic size asymmetry augments 
the colloidal charge neutralization and the screening in comparison with 
the size-symmetric case even at high values of $\sigma_0$, 
{\it proving the non-dominance of the counterions in the primitive model}. 
On the theoretical side, the HNC/HNC and HNC/MSA integral equations displayed consistent 
results with MC simulations, without a notable predominance in the accuracy between them, whereas MGC and 
URMGC evidenced the limitations of the Poisson-Boltzmann theories. 
Other consequences of the ionic size asymmetry in the EDL that have been confirmed 
by the present numerical simulations and HNC/HNC and HNC/MSA calculations 
are the appearance of charge reversal in monovalent 
salts and the reentrance of the mean electrostatic potential at the outer Helmholtz plane 
(i.e., the change of sign of the $\psi_{OHP}$ from negative to positive and, then, to 
negative again when $\sigma_0$ increases from zero) for divalent salts. 
If the usual identification between the well-known electrokinetic zeta 
potential and the mean electrostatic potential in the neighborhood 
of the Helmholtz region is assumed \cite{Hu01}, 
such reentrance in $\psi_{OHP}$ could be of relevance for mobility experiments 
since it indicates the possibility of observing 
an alternating direction in the electrophoresis of a colloid, immersed in a multivalent 
electrolytic bath, as a function of the native macroparticle's charge. 
Furthermore, the reported MC simulations have also evinced that anomalous curvatures 
can appear at the OHP in the primitive model EDL, which could be important for several 
recent investigations about non-intuitive phenomena (e.g., the appearance of negative 
differential capacitances \cite{Att01,To01,Par01,Par02,Ki01,Par03}) in 
electrolyte-electrode systems. 
In summary, the data reported in this paper suggest that the ionic size and, especially, 
the ionic size asymmetry should be considered as very sensitive parameters 
that, in combination with the concentration and valence of the electrolyte and the macroion's surface 
charge density, control the electrostatic-entropic coupling in the primitive model EDL. 
In particular, given that for multivalent salts the ionic hydration 
augments notably the finite size and size asymmetry effects, this could 
represent a way to attain a high electrostatic-entropic coupling at reasonable 
experimental conditions for which the charge reversal could 
be detected \cite{Mar01,Bes01,Bes02,Fer01,Hey01}. 

\begin{acknowledgments}
This work was supported by Consejo Nacional de Ciencia y Tecnolog\'{\i}a (CONACYT, M\'exico), 
through grants CB-2006-01/58470 and C01-47611, and PROMEP. E. G.-T. 
thanks to Centro Nacional de Superc\'omputo of Instituto Potosino de
Investigaci\'on Cient\'{\i}fica y Tecnol\'ogica for the computing time in the Cray XD1
and IBM E-1350 machines. 
G. I. G.-G. acknowledges posdoctoral fellowship from CONACYT. 
\end{acknowledgments}

\end{document}